\documentclass{article}

\usepackage{arxiv}

\usepackage[utf8]{inputenc} 
\usepackage[T1]{fontenc}    
\usepackage{hyperref}       
\usepackage{url}            
\usepackage{booktabs}       
\usepackage{amsfonts}       
\usepackage{nicefrac}       
\usepackage{microtype}      
\usepackage{lipsum}		
\usepackage{graphicx}
\usepackage{natbib}
\usepackage{doi}

\title{Exploring the Use of Machine Learning Weather Models in Data Assimilation}

\author{ \href{https://orcid.org/0000-0002-2832-4790}{\includegraphics[scale=0.06]{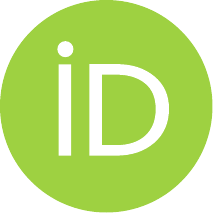}\hspace{1mm}Xiaoxu Tian}\thanks{Corresponding author: Xiaoxu Tian, xtian15@terpmail.umd.edu}  \\
	NOAA/NCEP/EMC\\
	College Park, MD 20740 \\
	\And
	Daniel Holdaway \\
	NOAA/NCEP/EMC\\
	College Park, MD 20740 \\
	\And
	Daryl Kleist \\
	NOAA/NCEP/EMC\\
	College Park, MD 20740 \\
}

\date{}

\renewcommand{\shorttitle}{ML Applications in DA}

\hypersetup{
pdftitle={ML Weather Models in Data Assimilation},
pdfsubject={ML and Data Assimilation},
pdfauthor={Xiaoxu Tian, Daniel Holdaway, and Daryl Kleist},
pdfkeywords={ML weather models, data assimilation, weather forecast},
}

\begin{document}
\maketitle

\begin{abstract}
The use of machine learning (ML) models in meteorology has attracted significant attention for their potential to improve weather forecasting efficiency and accuracy. GraphCast and NeuralGCM, two promising ML-based weather models, are at the forefront of this innovation. However, their suitability for data assimilation (DA) systems, particularly for four-dimensional variational (4DVar) DA, remains under-explored. This study evaluates the tangent linear (TL) and adjoint (AD) models of both GraphCast and NeuralGCM to assess their viability for integration into a DA framework.

We compare the TL/AD results of GraphCast and NeuralGCM with those of the Model for Prediction Across Scales - Atmosphere (MPAS-A), a well-established numerical weather prediction (NWP) model. The comparison focuses on the physical consistency and reliability of TL/AD responses to perturbations. While the adjoint results of both GraphCast and NeuralGCM show some similarity to those of MPAS-A, they also exhibit unphysical noise at various vertical levels, raising concerns about their robustness for operational DA systems.

The implications of this study extend beyond 4DVar applications. Unphysical behavior and noise in ML-derived TL/AD models could lead to inaccurate error covariances and unreliable ensemble forecasts, potentially degrading the overall performance of ensemble-based DA systems, as well. Addressing these challenges is critical to ensuring that ML models, such as GraphCast and NeuralGCM, can be effectively integrated into operational DA systems, paving the way for more accurate and efficient weather predictions.
\end{abstract}


\section{Introduction}
The field of meteorology has long relied on numerical weather prediction (NWP) models to provide accurate and timely weather forecasts. These models, based on the fundamental physical laws governing atmospheric dynamics, have become increasingly sophisticated over the decades \citep{RN323}. However, the advent of machine learning (ML) has introduced new opportunities for enhancing weather forecasting efficiency and accuracy, offering a complementary approach to traditional NWP models \citep{RN324,RN325,RN326}. This study focuses on the application of ML in data assimilation (DA) systems, specifically evaluating the suitability of an ML-based weather model, GraphCast, for integration into a four-dimensional variational (4DVar) DA framework. \\

NWP models, such as the Model for Prediction Across Scales - Atmosphere (MPAS-A), have demonstrated significant success in predicting weather patterns \citep{RN211,RN129}. These models solve the equations of motion numerically, incorporating various physical processes like radiation, convection, and microphysics. Despite their robustness, NWP models face challenges such as computational expense and the need for continuous improvements in parameterization schemes \citep{RN327}.\\

ML, particularly deep learning, has emerged as a powerful tool in many scientific domains due to its ability to learn complex patterns from large existing datasets. In meteorology, ML models have been employed for tasks ranging from short-term forecasting to climate projections \citep{RN328}. The potential for ML models to augment traditional NWP models lies in their ability to process vast amounts of data quickly and capture nonlinear relationships that might be challenging for conventional models \citep{RN324,RN325,RN326}. GraphCast is an example of an ML-based weather model designed to leverage graph neural networks for forecasting. This model has shown promise in initial studies, demonstrating the capability to produce competitive weather forecasts \citep{RN324}. \cite{RN326} introduced NeuralGCM, a hybrid weather and climate model that combines a traditional dynamical core with neural network-based parameterizations of physical processes. This approach uses machine learning to improve the representation of subgrid-scale phenomena, offering potential advancements in forecasting accuracy and computational efficiency. However, the application of GraphCast and NeuralGCM in data assimilation systems, particularly with 4DVar algorithms, remains largely unexplored.\\

Data assimilation is a critical component of modern weather forecasting systems, combining observational data with model forecasts to produce the best possible estimate of the current state of the atmosphere. The 4DVar DA system is one of the most advanced techniques, optimizing the initial conditions of a weather model by minimizing the difference between the model forecast and observational data over a specified time window \citep{RN48,RN50,RN51}. The success of a 4DVar DA system relies on the accuracy of the tangent linear (TL) and adjoint (AD) models, which are used to propagate perturbations forward and sensitivities backward in time, respectively. These models need to be physically consistent and accurate to ensure the DA systems can effectively correct model states based on observational data \citep{RN248}.\\

Several studies have explored the integration of ML models into various aspects of weather forecasting. For instance, \cite{RN271} highlighted the advantages of using ML models for emulating subgrid-scale processes in NWP models. \cite{RN307} investigated the possibility of applying the adjoint of a neural network emulator in a 4DVar DA framework. While ML models like GraphCast hold promise, integrating them into DA systems presents significant challenges. Unlike traditional NWP models, ML models do not inherently adhere to physical laws unless explicitly designed to do so \citep{RN329}. This can lead to issues with physical consistency, particularly in the TL and AD models, which are crucial for the 4DVar DA system. Initial investigations into the TL/AD models in this study have revealed several unphysical properties, raising concerns about their suitability for data assimilation, especially 4DVar applications. These unphysical behaviors can also lead to inaccurate error covariances and unreliable ensemble forecasts, undermining the overall effectiveness of the DA system \citep{RN330}. Therefore, a thorough evaluation of the TL and AD models of both GraphCast and NeuralGCM is necessary to determine their viability for integration into a 4DVar DA framework.\\

The primary objective of this study is to assess the suitability of TL and AD of ML-based models (GraphCast and NeuralGCM) for 4DVar DA systems. The aim is to compare the TL and AD models of both GraphCast and NeuralGCM with those of the MPAS-A model, focusing on their physical consistency and reliability in response to perturbations \citep{RN227,RN269,RN241}. By identifying the unphysical behaviors of ML models' TL and AD, this study aims to highlight potential issues when applying ML models in DA systems, thereby encouraging the development of better strategies for their integration. The authors hope to uncover underlying challenges that may arise in both ensemble-based and variational DA methods, ultimately contributing to more effective incorporation of ML models in operational DA systems and enhancing overall weather forecasting accuracy and efficiency.\\

\section{Methods}
The first machine learning model of choice in this study is GraphCast, a graph neural network (GNN)-based model for weather forecasting \citep{RN306,RN324}. GNNs are particularly suitable for this task because they can naturally handle the irregular grid structures commonly found in atmospheric models, allowing for efficient processing of spatial data. The GNN is composed of nodes and edges, where nodes represent the grid points, and edges represent the connections between them. The state of the atmosphere at each node is represented by a feature vector \(\mathbf{h}_i\). The update rule for the GNN can be generally written as:
\[ \mathbf{h}_i^{(k+1)} = \sigma \left( \sum_{j \in \mathcal{N}(i)} \mathbf{W} \mathbf{h}_j^{(k)} + \mathbf{b} \right) \]
where \(\mathbf{h}_i^{(k)}\) is the feature vector of node \(i\) at layer \(k\), \(\mathcal{N}(i)\) represents the neighbors of node \(i\), \(\mathbf{W}\) is the weight matrix, \(\mathbf{b}\) is the bias term, and \(\sigma\) is the activation function. To develop the TLM and adjoint models for GraphCast, we linearize the GNN update rules around the current state. For the TLM, this involves computing the Jacobian matrix of the GNN with respect to the input features. The adjoint model is then obtained by transposing the Jacobian matrix. \\

The second machine learning model used in this study is NeuralGCM, a hybrid weather and climate model that integrates a traditional dynamical core for solving the discretized version of the governing dynamical equations with neural network-based parameterizations of unresolved physical processes \citep{RN326}. NeuralGCM enhances the strengths of machine learning to improve the representation of complex, subgrid-scale processes like cloud formation, radiative transport, precipitation and subgrid-scale dynamics, which are challenging to simulate using traditional methods. The model's physical parameterizations are represented by deep neural networks (DNNs) that take various atmospheric state variables as input. Mathematically, the neural network-based parameterization in NeuralGCM can be expressed as:
\[
 \mathbf{y} = \phi(\mathbf{x}; \theta) 
\]
 
where $\mathbf{x}$ represents the atmospheric input variables, $\mathbf{y}$ represents the parameterized outputs, and $\theta$ denotes the learnable parameters of the neural network. To develop the tangent linear and adjoint models of NeuralGCM, the neural networks are linearized around the current state, and the gradients with respect to the model’s inputs are computed using automatic differentiation. The adjoint model is then constructed by computing the transpose of the Jacobian of the neural network’s output with respect to the inputs.

\subsection{Tangent Linear and Adjoint (TL/AD) Models}
The tangent linear model (TLM) represents the linearized version of the nonlinear forecast model around a given state. It is essential for data assimilation, particularly in variational methods like 4DVar, as it provides the linear relationship between small perturbations in the initial conditions and the resulting changes in the forecast.

Given a nonlinear model \(\mathcal{M}\) that maps an initial state \(\mathbf{x}_0\) to a forecast state \(\mathbf{x}_t\) after time \(t\):
\[ \mathbf{x}_t = \mathcal{M}(\mathbf{x}_0) \]

The TLM, \(\mathbf{M}\), of \(\mathcal{M}\) at \(\mathbf{x}_0\) is defined as:
\[ \delta \mathbf{x}_t = \mathbf{M}(\mathbf{x}_0) \delta \mathbf{x}_0 \]

Here, \(\delta \mathbf{x}_0\) represents a small perturbation in the initial state, and \(\delta \mathbf{x}_t\) is the corresponding change in the forecast state.\\

The adjoint model, \(\mathbf{M}^*\), is defined as the transpose (or adjoint) of the tangent linear model:
\[ \mathbf{M}^* = \mathbf{M}^T \]

For a given cost function \(J\) that depends on the forecast state \(\mathbf{x}_t\), the gradient of \(J\) with respect to the initial state \(\mathbf{x}_0\) can be computed using the adjoint model. \\

\subsection{Verification of TLM and Adjoint Model Correctness}
The correctness of the TLM can be verified by checking that the model correctly approximates the nonlinear model for small perturbations. This involves ensuring that the finite difference approximation of the nonlinear model matches the TLM output:
\[
\Psi (\alpha)=
\frac{\| \mathcal{M}(\mathbf{x} + \alpha \mathbf{p}) 
- \mathcal{M}(\mathbf{x}) \|}
{\| \mathbf{M}(\mathbf{x}) \alpha \mathbf{p} \|} = 1 + O(\alpha)
\]
for sufficiently small \(\alpha\). The verification results for GraphCast TLM are shown in Fig. 1. The top panel shows the variation of \(\Psi (\alpha) \) with respect to \(\alpha\). The bottom two panels show the relationship between the numerator and denominator for temperature and specific humidity at different grid points when \(\alpha = 10^{-4}\), which proves quite close to the \(y=x\) relationship.\\

The adjoint model can then be verified by ensuring it satisfies the adjoint consistency property. This involves checking that the inner product of the TLM with any perturbation \(\delta \mathbf{x}_0\) matches the inner product of the adjoint model with the corresponding adjoint state \(\mathbf{\lambda}\):
\[ \left\langle \mathbf{M}(\mathbf{x}_0) \delta \mathbf{x}_0, \mathbf{\lambda} \right\rangle = \left\langle \delta \mathbf{x}_0, \mathbf{M}^*(\mathbf{x}_0) \mathbf{\lambda} \right\rangle \]
for any random vectors \(\delta \mathbf{x}_0\) and \(\mathbf{\lambda}\). In the case of this study calculating with double precision, the left-hand side of the equation above is 2679547284.519031, the right-hand side is 2679547284.519030, with 15 digits in agreement, close to the theoretical limit.

\subsection{TL/AD Comparison of GraphCast, NeuralGCM, and MPAS-A}
To assess the validity of the GraphCast and NeuralGCM TL/AD models, we study a number of case studies. In determining the potential usefulness of an adjoint, the structures of the model fields are evaluated for meaningful physical interpretation. To help with this assessment a physical based adjoint is included in the comparison. For this physical model the MPAS-A system is chosen. MPAS-A has adjoint and tangent linear versions of the dynamical core \citep{RN227}. It should be noted that all these models are whole atmosphere prediction systems i.e. modeling both resolved (in terms of the discrete grid spacing) and unresolved processes, which represent processes like convection and radiation. For GraphCast and NeuralGCM the adjoint of the model inherently includes all these unresolved processes as well. However the adjoint of MPAS-A does not. Since these unresolved processes are modeled using highly nonlinear physical models, writing their adjoint is non-trivial. So the linearized version of MPAS-A only describes the evolution of resolved scale processes, which tend to be more linear at the resolutions being considered. While lacking the adjoint of unresolved processes it is still very useful to have the MPAS-A adjoint since it will provide a benchmark. Further some of the test cases considered are chosen so as to minimize the impact of the unresolved processes. \\

The comparison of the three systems will involve analyzing the TL responses to perturbations in initial conditions and the adjoint sensitivities to a response function. Key metrics for comparison will include the spatial distribution of perturbations, as well as the consistency of adjoint sensitivity with respect to a defined response function. This comprehensive evaluation will help identify any unphysical behaviors or inconsistencies in the TL/AD models of GraphCast and NeuralGCM, determining their suitability for integration into data assimilation frameworks.

 \section{Results and Comparisons}
The performance and validity of the GraphCast TL/AD models developed in this study are assessed through a series of comparisons with the TL/AD models of the MPAS-A. This section presents the results of these comparisons, highlighting key similarities and differences in their responses to initial perturbations.\\

The initial conditions for both GraphCast and MPAS-A models are from the ERA5 reanalysis data at 00 UTC on January 1, 2022. The top panel of Figure 2 provides the context of the background state used for this analysis. The perturbation is imposed on the zonal wind variable only at one spot, location of which is chosen at the core of a jet stream at 250 hPa, characterized by strong zonal wind speeds as shaded. This specific location is selected due to the dynamic nature of the jet stream, making it an ideal region to test the sensitivity and accuracy of the TL/AD models. The middle and bottom rows of Figure 2 illustrate the horizontal distribution and vertical cross-sections of the TL responses to a zonal wind perturbation imposed at the marked location. The left column displays the results from the GraphCast TLM, the right column those from the MPAS-A TLM. \\

In the horizontal distribution plots, the GraphCast TLM shows a clear response to the perturbation, with noticeable changes in the zonal wind field extending downstream of the perturbation location. This downstream feature, captured in both GraphCast and MPAS-A, suggests the presence of a Rossby mode in the dynamics of both models. However, the propagation distance and magnitudes of the Rossby mode differ significantly between the two cases after the same 6-hour TLM forecast. The most striking difference is the strong perturbation value found precisely at the original perturbation spot in the case of GraphCast, even after 6 hours of propagation. This feature does not appear to be physical, as no such signal is present in the case of MPAS-A. \\

The vertical cross-sections provide further insights into the model behaviors. Both GraphCast and MPAS-A display downstream responses at the jet core, although with opposite signs. The most concerning signal is the strong TL response at the original perturbation spot in GraphCast, which extends both upward and downward. This persistent and localized response suggests potential issues with how GraphCast TLM handles the propagation of initial perturbations, raising questions about the physical realism. \\

The horizontal distribution of adjoint sensitivity 6 hours prior to the response function \( T_{ad} = 1 \) at 06 UTC on January 1, 2022, at the perturbation location marked by the cross is shown in Figure 3. The top row presents the adjoint sensitivity in temperature, the middle row in zonal wind, and the bottom row in specific humidity. The left column shows the GraphCast adjoint results, while the right column presents the MPAS-A adjoint results. \\

In the GraphCast adjoint model, there is no sensitivity in temperature 6 hours prior to the response function, except at the original perturbation spot. This result contrasts with the MPAS-A adjoint, which shows upstream sensitivity following a Rossby mode pattern. The absence of such a pattern in GraphCast suggests unphysical behavior, as the adjoint sensitivity should typically exhibit upstream features due to the nature of Rossby wave propagation. In the field of zonal wind, the GraphCast adjoint model shows upstream sensitivity with patterns and magnitudes similar to those in MPAS-A. However, the GraphCast adjoint results appear slightly noisier compared to the smoother patterns in MPAS-A. Additionally, the GraphCast model exhibits strong sensitivity precisely at the original perturbation location, which does not appear to be physically realistic. In the field of specific humidity shown at the bottom row, the GraphCast adjoint model shows upstream sensitivity patterns with comparable propagation distances. However, the magnitudes of the GraphCast sensitivities are an order of magnitude stronger than those in MPAS-A, raising concerns about their physical realism. Furthermore, the GraphCast model also displays strong sensitivity at the perturbation location, which is not physically realistic. \\

The vertical cross-sections of the same adjoint sensitivity as in Figure 3 are shown in Figure 4. Vertically, there is no upstream sensitivity in temperature, except for the strong on-the-spot feature at the original perturbation location. This result contrasts sharply with the MPAS-A adjoint, which shows a more physically realistic upstream sensitivity pattern. The vertical distribution of zonal wind sensitivity in the GraphCast adjoint model is quite noisy, with significant signals appearing at various vertical levels. This diffuse pattern is concerning and suggests unrealistic behavior, which would be problematic if applied in a data assimilation system. In contrast, the MPAS-A adjoint model shows a more coherent and physically consistent vertical distribution of sensitivity. In the field of specific humidity, the GraphCast adjoint model shows sensitivity signals with even greater magnitudes at various vertical levels, primarily above 250 hPa. These strong sensitivities do not have a corresponding pattern in the MPAS-A results, raising concerns about the physical realism of the GraphCast adjoint model. \\

In addition to the findings from GraphCast and MPAS-A, the NeuralGCM adjoint model exhibits some promising physical features similar to those seen in the MPAS-A adjoint, such as identifiable upstream sensitivity in key variables (Fig. 5), given that the a traditional dynamical core is used in NeuralGCM. However, a noticeable issue with NeuralGCM is the presence of noise at various vertical levels, particularly in the temperature field. This noise is likely attributed to challenges in the adjoint calculated from the machine learning-based physical parameterizations, where the neural network’s approximations introduce unphysical signals. While NeuralGCM shows potential in capturing important sensitivities, the noisy results at specific vertical layers suggest that further refinement is required to achieve the level of smoothness and physical realism demonstrated by the MPAS-A adjoint. \\

These observations highlight critical issues in the GraphCast adjoint model that need to be addressed to improve its physical realism and applicability in data assimilation systems. The unrealistic on-the-spot sensitivity and the diffuse and noisy patterns in the zonal wind and specific humidity fields suggest that further refinement of the GraphCast adjoint model is necessary to ensure reliable performance in operational settings. \\

\section{Discussion and Conclusions}

The application of machine learning (ML) models in meteorology has gained significant attention due to their potential to enhance both forecasting efficiency and accuracy. Among these, GraphCast and NeuralGCM present promising innovations in ML-based weather modeling. However, the suitability of such models for data assimilation (DA) systems, particularly within a four-dimensional variational (4DVar) framework, demands a thorough evaluation. This study develops and investigates the tangent linear (TL) and adjoint (AD) models of both GraphCast and NeuralGCM to assess their viability for integration into 4DVar DA. We compare the results from these ML models to those of MPAS-A, a traditional NWP model known for its robust adjoint and tangent linear models. \\

The GraphCast TLM shows clear responses to initial perturbations with changes in the zonal wind field that extend downstream, suggesting the presence of Rossby wave dynamics. However, differences between GraphCast and MPAS-A are evident in the propagation distance and magnitudes of these waves, with GraphCast showing a strong and persistent perturbation at the original spot, even after 6 hours of propagation. This localized sensitivity is unphysical and absent in MPAS-A, raising concerns about the realism of GraphCast’s adjoint behavior in handling perturbation propagation. \\

Similarly, NeuralGCM exhibits some physical features in its adjoint sensitivities, such as identifiable upstream propagation, but also displays notable noise at various vertical levels. This noise likely stems from the adjoint of ML-based physical parameterizations, where neural network approximations may introduce unphysical signals. Despite showing promise in capturing important adjoint sensitivities, the NeuralGCM adjoint model’s noisier results, suggest that further refinements are needed to achieve the same level of smoothness and physical consistency as the MPAS-A adjoint. \\

The adjoint sensitivity analysis of both GraphCast and NeuralGCM underscores the differences when compared to physical NWP models like MPAS-A. For GraphCast, the temperature field shows no upstream sensitivity beyond the strong on-the-spot feature, contrasting sharply with MPAS-A’s upstream Rossby mode. NeuralGCM displays promising and physical upstream patterns in adjoint sensitivity fields, accompanied by noisy signals across vertical levels. While both GraphCast and NeuralGCM demonstrate encouraging capabilities in handling some essential atmospheric processes, such as Rossby wave dynamics, their current formulations present challenges for integration into DA systems. The strong localized sensitivities and noisy patterns suggest that further model refinement is necessary to improve their physical realism and applicability in operational data assimilation. \\

The findings of this study indicate that while machine learning-based models like GraphCast and NeuralGCM hold promise for advancing weather prediction, their current tangent linear and adjoint models exhibit several unphysical properties that limit their suitability for data assimilation applications. The persistent strong sensitivities at the perturbation spot, noisy adjoint patterns, and exaggerated magnitudes, especially in specific humidity, suggest that further refinements are necessary before these models can be considered reliable for operational DA systems. Consider, for example, if a single observation was assimilated in a 4DVar system that utilized the adjoint of GraphCast. The resulting increment would be noisy and with structures unphysically far away from the observation. Applying that increment may well improve the fit to that observation but since the structures are so unphysical it would disrupt the solution elsewhere and may increase the overall forecast error as the simulation proceeds. \\

These issues are not limited to 4DVar variational DA systems. They also imply potential challenges in Ensemble Kalman Filter (EnKF) DA methods, as ensemble-based methods rely on perturbing the model state to generate error covariances. The unrealistic sensitivity to perturbations and the noisy adjoint patterns point to distorted error covariances, which would degrade the quality of state estimation in an EnKF system. The overestimated or misplaced sensitivities would likely result in poor representation of background error covariances, ultimately affecting the assimilation of observations and leading to unreliable forecasts. Therefore, the tangent linear and adjoint issues identified in this study highlight the broader need for improved physical realism and numerical stability in these ML models before they can be confidently applied across different DA methods, including EnKF. \\

\section*{Acknowledgments}
The authors would like to thank DeepMind for publishing their GraphCast model and the benchmark data, which have been invaluable for this study.

\bibliographystyle{unsrtnat}
\bibliography{references}  

\newpage
\begin{figure}
\noindent\includegraphics[width=\textwidth]{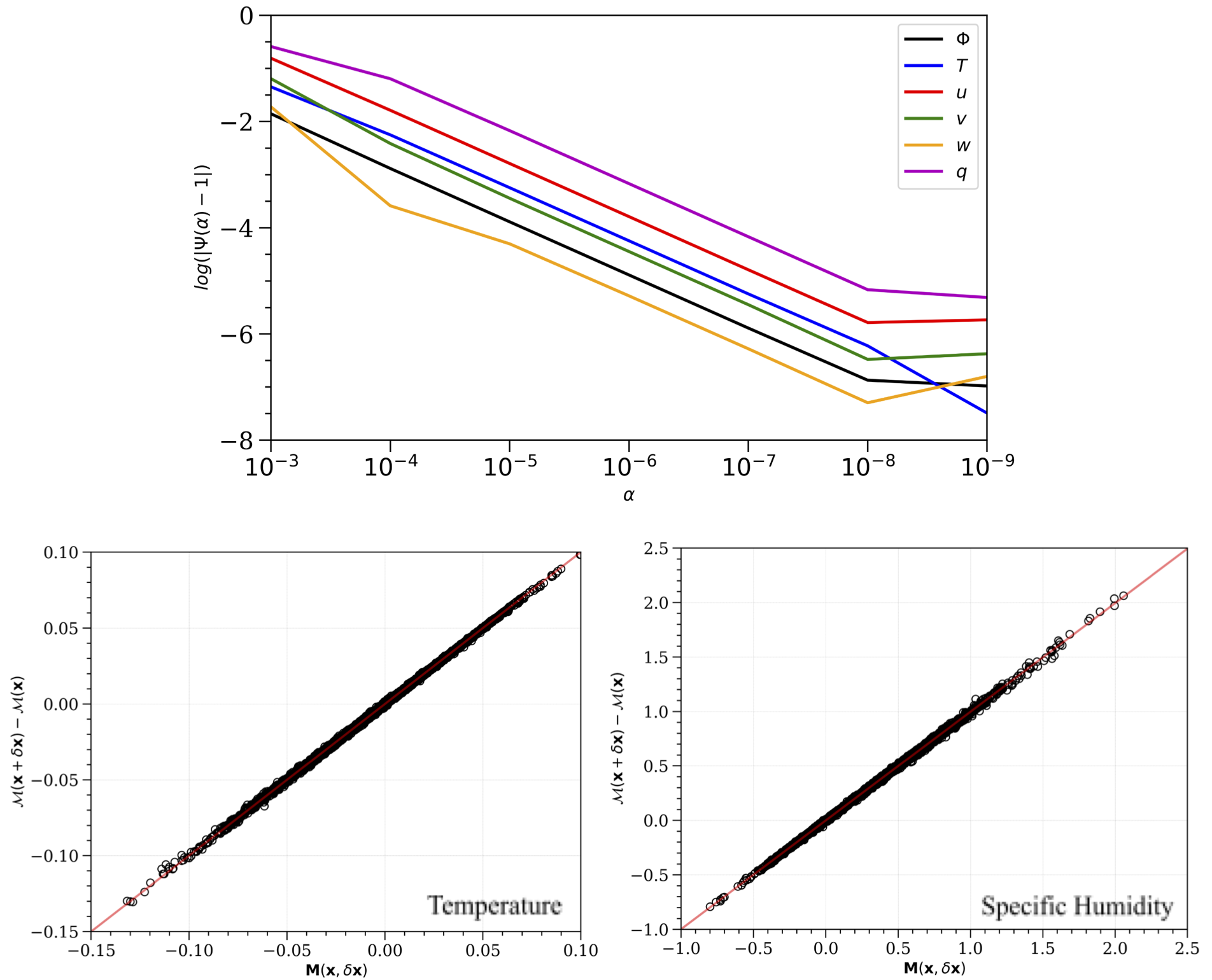}
\caption{Variations in the function for the correctness check of the GraphCast tangent linear model for a 6-hour forecast length when the initial conditions for variables $\Phi$, $T$, $u$, $v$, $w$, and $q$ are separately perturbed, where $\alpha$ is the scale factor of initial perturbations. (a) Logarithm of the absolute value of $\frac{J(\alpha)}{\alpha} - 1$ for each variable, indicating the accuracy of the linear approximation. (b)-(c) Scatter plots of the nonlinear forward difference vs. tangent linear results for (b) temperature and (c) specific humidity. The slope/intercept of the linear regression line, plotted in red, are (b) 1.0/1.18×10-6 and (c) 0.997/2.31×10-9, respectively.
}
\end{figure}

\newpage
\begin{figure}
\noindent\includegraphics[width=\textwidth]{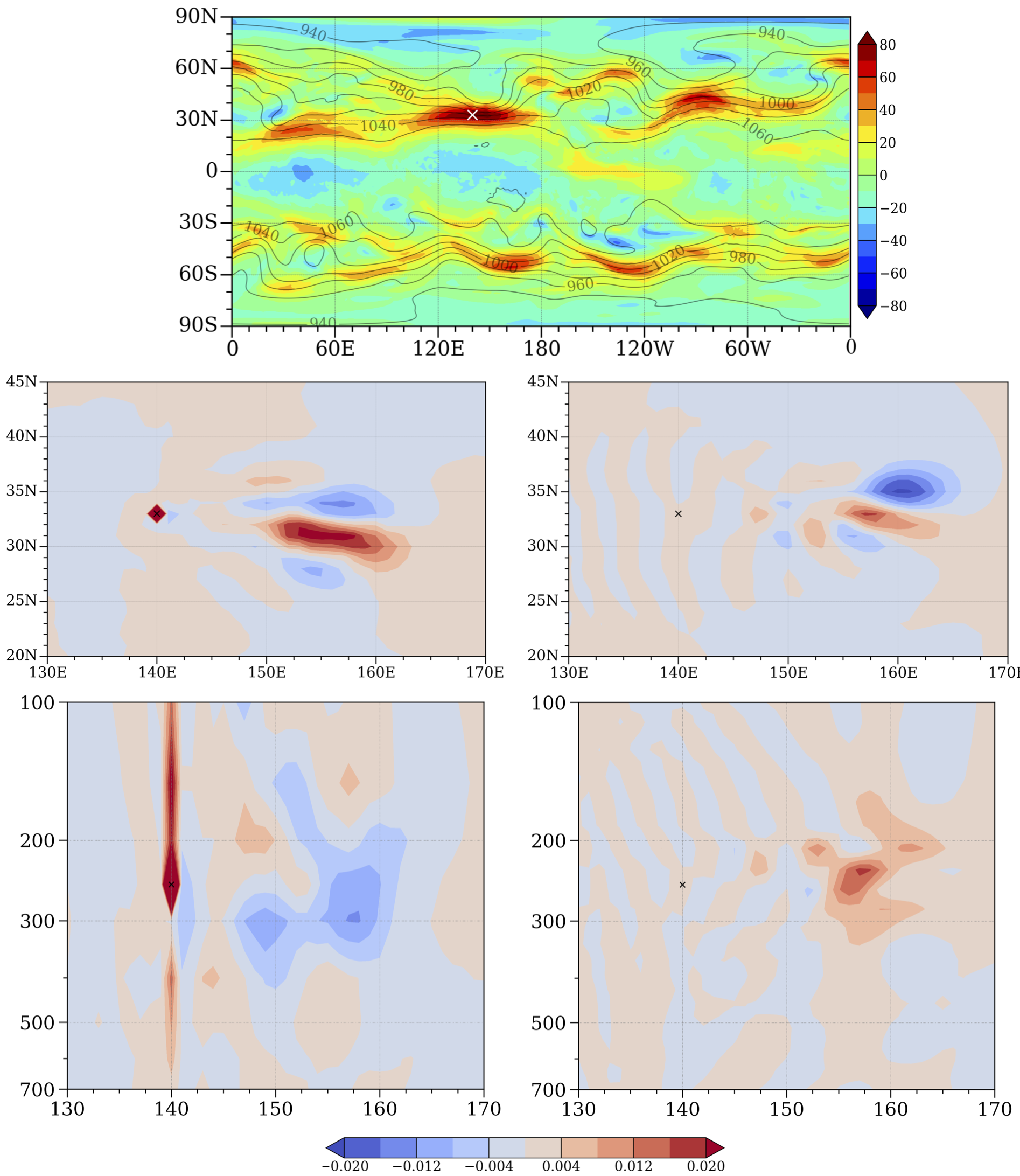}
\caption{(a) Background geopotential heights (contoured) and zonal wind (shaded) at 00 UTC on January 1, 2022. The cross marks the location of the imposed perturbation. (b)-(c) Horizontal distribution of the TL response in zonal wind 6 hours into the forecast to a zonal wind perturbation at the initial time for GraphCast (left) and MPAS-A (right). (d)-(e) Vertical cross-sections of the TL response in zonal wind along the longitude line at 33$^{\circ}$N for GraphCast (left) and MPAS-A (right).
}
\end{figure}

\newpage
\begin{figure}
\noindent\includegraphics[width=\textwidth]{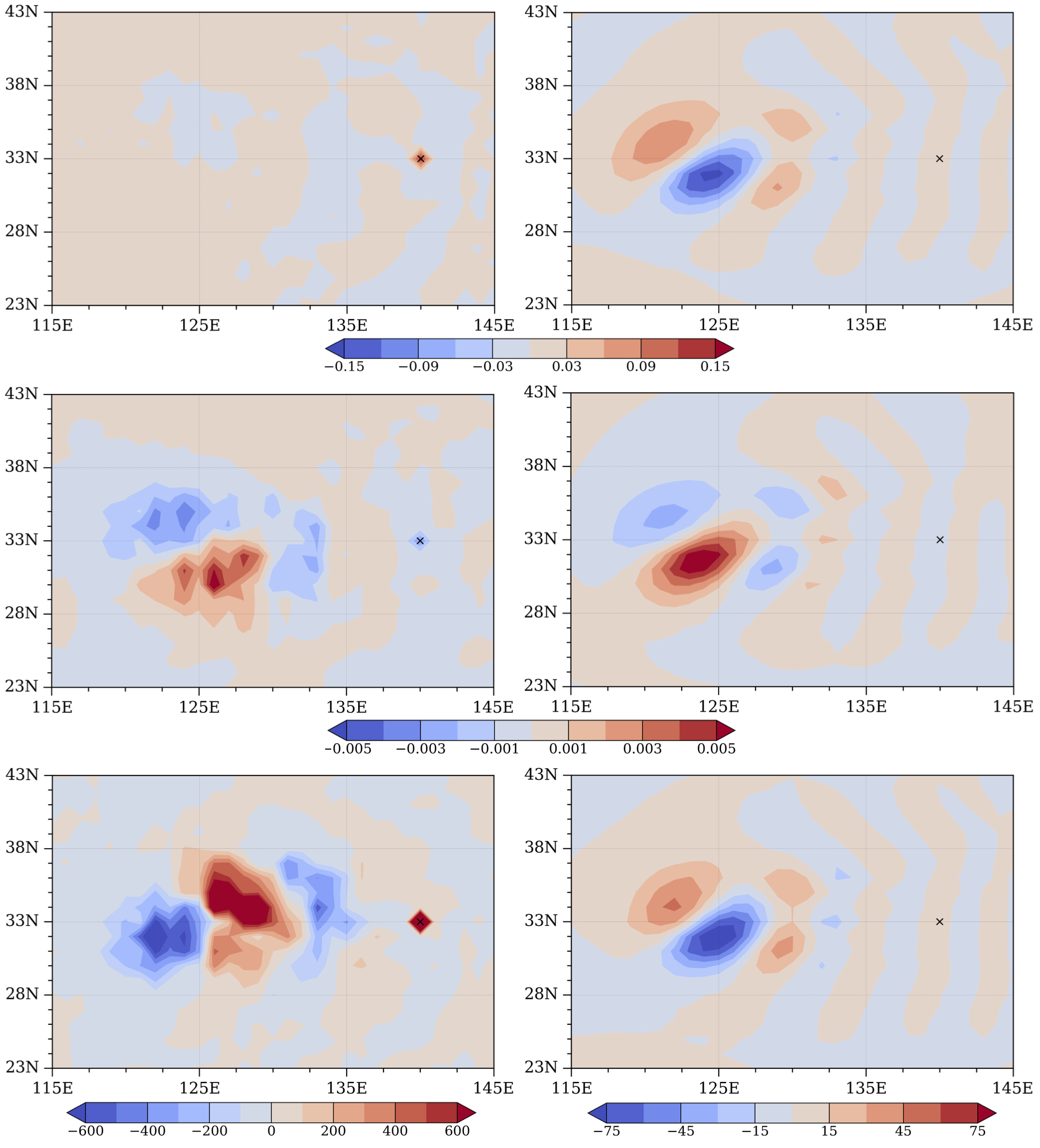}
\caption{Horizontal distribution of adjoint sensitivity 6 hours prior to the response function of $T=1$ marked by the cross in temperature (top row), zonal wind (middle row), and specific humidity (bottom row). The left column presents the GraphCast adjoint results, while the right column presents the MPAS-A adjoint results. 
}
\end{figure}

\newpage
\begin{figure}
\noindent\includegraphics[width=\textwidth]{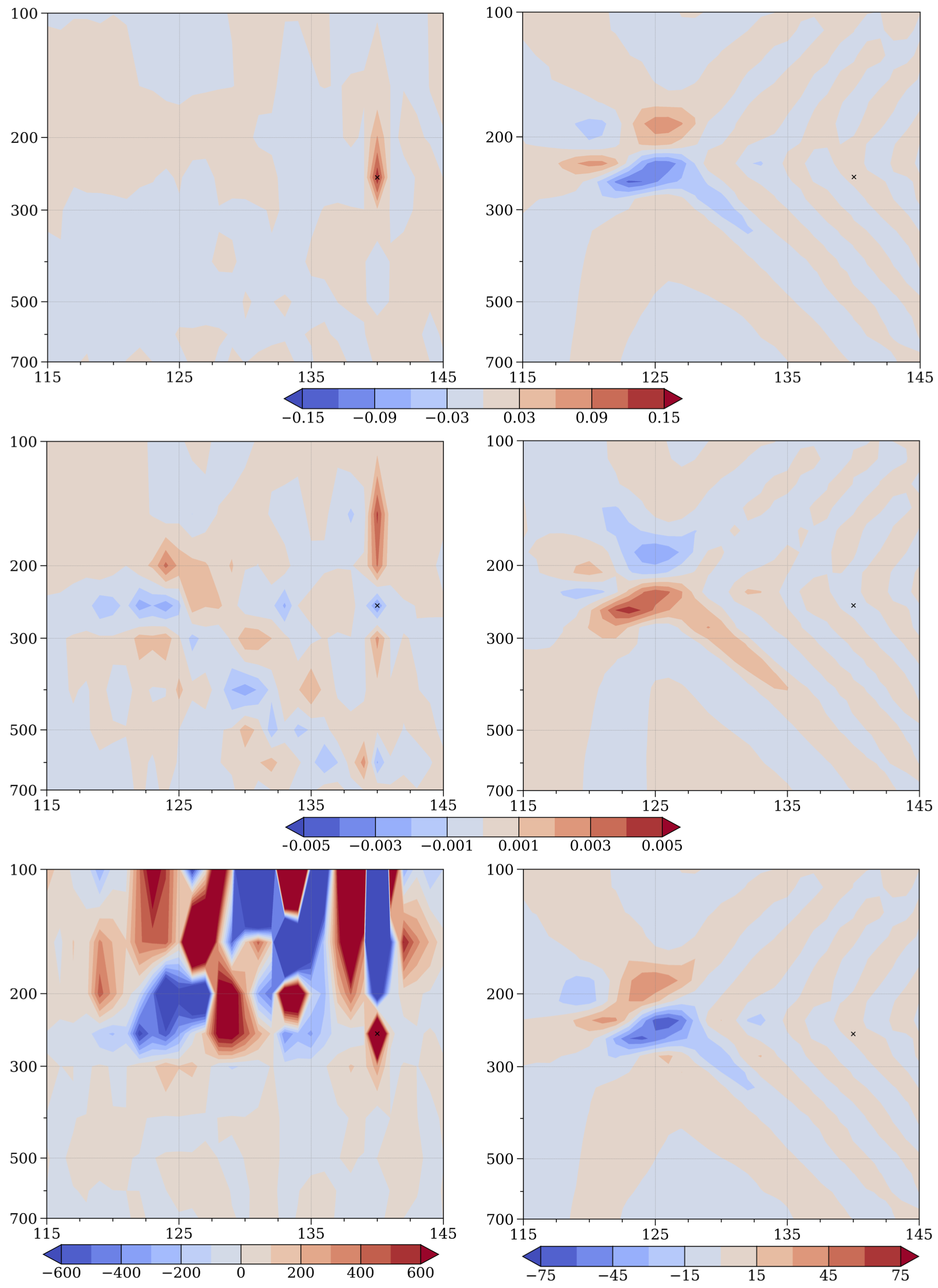}
\caption{Vertical cross-sections of adjoint sensitivity 6 hours prior to response function of $T_{ad}=1$ at the location marked by the cross along the longitude line at 33 degrees north temperature (top row), zonal wind (middle row), and specific humidity (bottom row). The left column presents the GraphCast adjoint results, while the right column presents the MPAS-A adjoint results.
}
\end{figure}

\newpage
\begin{figure}
\noindent\includegraphics[width=\textwidth]{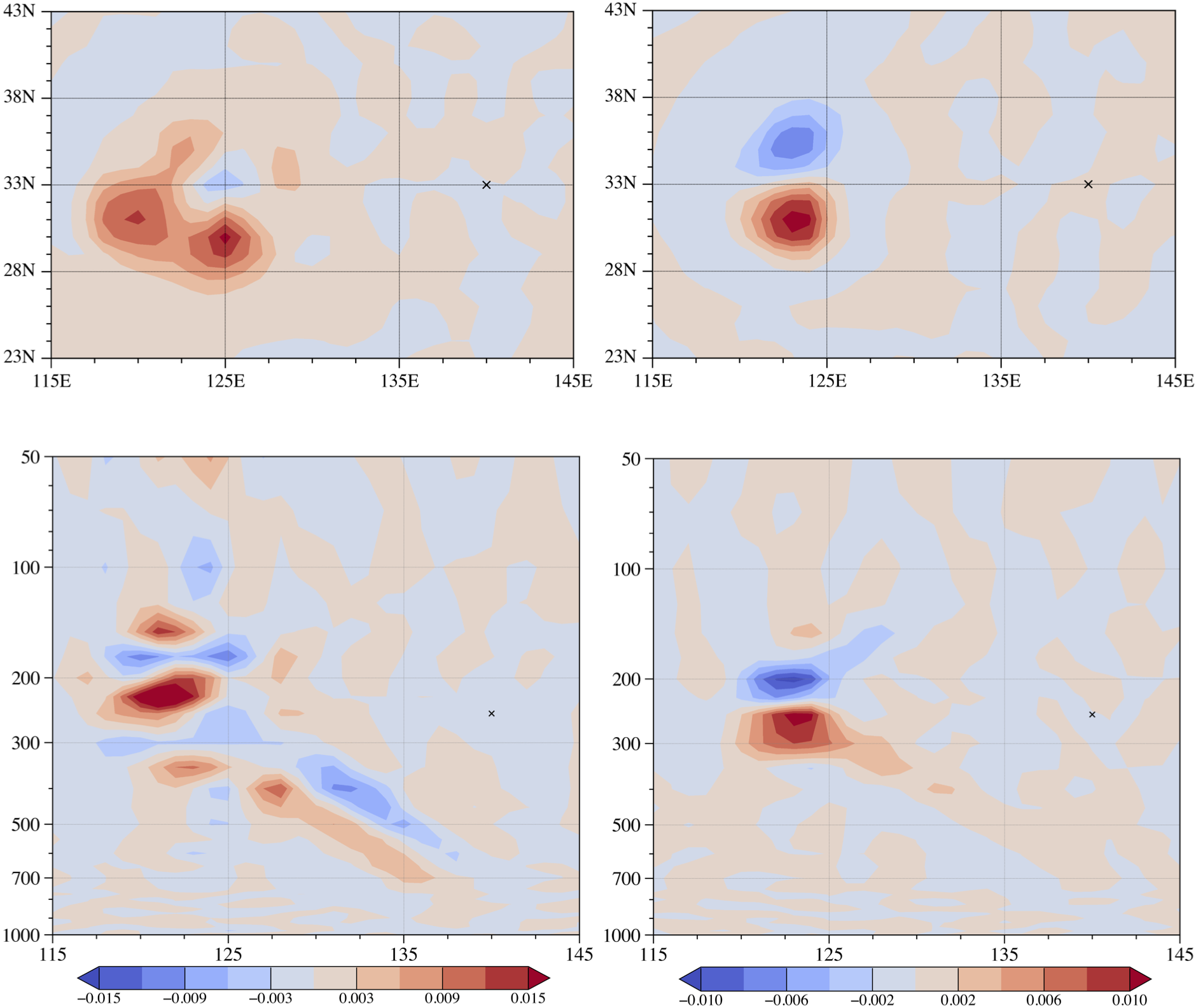}
\caption{The horizontal distribution (top row) and vertical cross sections (bottom row) of NeuralGCM adjoint sensitivity six hours prior to $T=1$ marked by the cross in temperature (left column) and zonal wind (right column).
}
\end{figure}






\end{document}